\input{aipcheck}

\documentclass[
    ,final            
  ]
  {aipproc}

\layoutstyle{8x11single}

\def\al{\alpha}

\def\th{\theta}

\def\si{\sigma}

\def\ta{\tau}

\def\om{\omega}

\def\De{\Delta}

\def\Om{\Omega}

\def\fr#1#2{\frac{#1}{#2}}

\def\beq{\begin{eqnarray}}
\def\eeq{\end{eqnarray}}

\begin{document}

\title{Meson Exchange Current (MEC) Models in Neutrino Interaction Generators}


\author{Teppei Katori}{
  address={Massachusetts Institute of Technology, Cambridge, MA}
}

\begin{abstract}
Understanding of the so-called {\it 2 particle-2 hole (2p-2h)} effect 
is an urgent program in neutrino interaction physics 
for current and future oscillation experiments. 
Such processes are believed to be responsible for the event excesses observed 
by recent neutrino experiments.  
The 2p-2h effect is dominated by the meson exchange current (MEC), 
and is accompanied by a 2-nucleon emission from the primary vertex, 
instead of a single nucleon emission from 
the charged-current quasi-elastic (CCQE) interaction. 

Current and future high resolution experiments 
can potentially nail down this effect. 
For this reason, there are world wide efforts to model 
and implement this process in neutrino interaction simulations. 
In these proceedings, I would like to describe how this channel is modeled in 
neutrino interaction generators. 
\end{abstract}

\maketitle


\section{Meson Exchange Current (MEC)}

The MEC is conceived to be responsible for the so-called ``dip region''. 
In electron scattering (for review, see~\cite{benhar}), 
inclusive electron scattering measures the out-going electron from the interaction. 
Since the beam has well-defined energy, 
an energy transfer $\om$ and a 3-momentum transfer $\vec{q}$ 
are reconstructed with high accuracy. 
The inclusive cross section is the differential cross section as a function of energy transfer. 
In this space, with moderate electron beam energy, one can see 2 bumps, representing 
quasi-elastic (QE) scattering peak and $\De$-resonance peak. 
These are often reasonably modeled, however, many theories fail to reproduce a dip between these 2 bumps. 
This is the dip region. 
By adding the MEC, 
some models successfully reproduce inclusive cross section data including dip region~\cite{nieves_eeprime}. 
The contribution of MEC is larger in the transverse response than the longitudinal response. 


Therefore, the importance of MEC is known from electron scattering data. 
MEC is an interaction involved in 2 nucleons, or 2-body current, 
and it is classified in ``2 particle-2 hole (2p-2h)'' effect. 
Here, a weak boson from the leptonic current is 
exchanged by a pair of nucleons (2-body current), 
and believed to lead to 2-nucleon emission. 
The importance of this process in neutrino interactions was 
first pointed out by Martini {\it et al.}~\cite{martini_first} 
shortly after the MiniBooNE experiment showed their CCQE double differential cross section~\cite{MB_ccqe}. 
Several groups successfully reproduces the MiniBooNE CCQE or 
neutral current elastic (NCEL) cross section data~\cite{MB_ncel} by adding the MEC in 
their models~\cite{martini_dd,nieves_first,bodek_tem,mosel_dd,donnelly_dd}. 
However other approaches appear to work equally well~\cite{butkevich,meucci}. 

To understand weather the MEC is the responsible process for enhancement of recent neutrino 
data~\cite{MB_ccqe,MB_ncel,MB_others,MINOS_ccqe,SB_ccincl}, 
we need to compare more kinematic details of the data with simulations. 
Here, we describe how this process is implemented in GENIE neutrino simulation generator~\cite{GENIE}.

Through this note, I assume charged-current interaction on carbon target, also, 
a neutrino scatters with a neutron-proton pair 
(hence outgoing particles from the primary vertex are one muon and two protons). 

\section{Modeling MEC Neutrino Interaction in GENIE}

\begin{figure}[t]
\includegraphics[height=0.3\textheight]{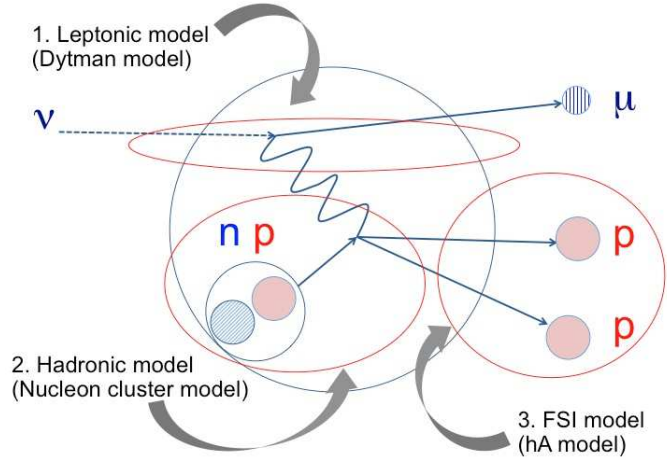}
\caption{\label{fig:mec}
Basic strategy of modeling MEC in GENIE.}
\end{figure}

Figure~\ref{fig:mec} shows a cartoon of how to model MEC in GENIE. 
Any interaction can be modeled with these three parts.

\begin{itemize}
\item Leptonic model
\item Hadronic model
\item Final state interaction (FSI) model
\end{itemize}

The leptonic model defines the differential cross section of the outgoing lepton 
for a given energy-momentum transfer. 
So this part completely specifies the lepton kinematics. 
There are several models available, 
and a neutrino interaction generator may choose one. 
Then using this energy-momentum transfer, 
the physics of the hadronic system needs to be specified. 
The problem is that there is no microscopic hadronic model of MEC is available. 
Distribution of initial nucleon kinematics, presence of correlations, type of nucleon pairs, etc, 
are key ingredients producing outgoing nucleons. 
This needs to be modeled. 
Finally, all outgoing particles experience FSI in the target nuclei. 
Observables always depend upon FSI, unfortunately. 

\subsection{Leptonic model, Dytman model}

To specify leptonic kinematics, GENIE collaboration developed the Dytman model. 
The Dytman model is motivated by the Lightbody model~\cite{lightbody}. 
In the Lightbody model, QE and $\De$ peaks are given by Gaussian distributions, 
then the MEC part is modeled as a Cauchy distribution between those 2 distributions.  
In the Dytman model, MEC is modeled as a Gaussian distribution between 2 peaks, QE and $\De$, 
to fill the dip region of electron scattering inclusive cross section data. 
The target nuclei dependence is linearly given. 

The differential cross sections for electron and neutrino scattering of the Dytman model are made 
using only the Sachs magnetic form factor ($G_{M,p/n}$), 
to emphasize the transverse nature of MEC~\cite{bodek_tem}. 
The Sachs magnetic form factor has power 6 (dipole form factor is power 2), 
motivated by the deuteron elastic scattering form factor. 
Here, electron-nucleon scattering can be written (where $\ta=\fr{Q^2}{4M^2}$), 
\beq
\fr{d^2\si_{p/n}}{d\Om dE'}=\fr{4\al^2{E'}^2}{Q^4}cos^2\left(\fr{\th}{2}\right)
\left[\fr{\ta}{1+\ta}+2\ta tan^2\left(\fr{\th}{2}\right)\right]G_{M,p/n}^2~,
\label{eq:electron}
\eeq
and the neutrino-nucleon CC cross section can be written, 
\beq
\fr{d\si_{p/n}}{dQ^2}=\fr{M^2{G_F}^2cos^2\th_c}{8\pi {E_{\nu}}^2}
\left\{\fr{m^2+Q^2}{M^2}\left[\ta-\fr{m^2}{4M^2}\right]+\frac{(s-u)^2}{M^4}\cdot\fr{\ta}{4(1+\ta)}\right\}G_{M,p/n}^2~.
\label{eq:neutrino}
\eeq
In this way, one can universally treat MEC in 
the electron scattering and the neutrino scattering processes.  

\begin{figure}[t]
\includegraphics[height=0.35\textheight]{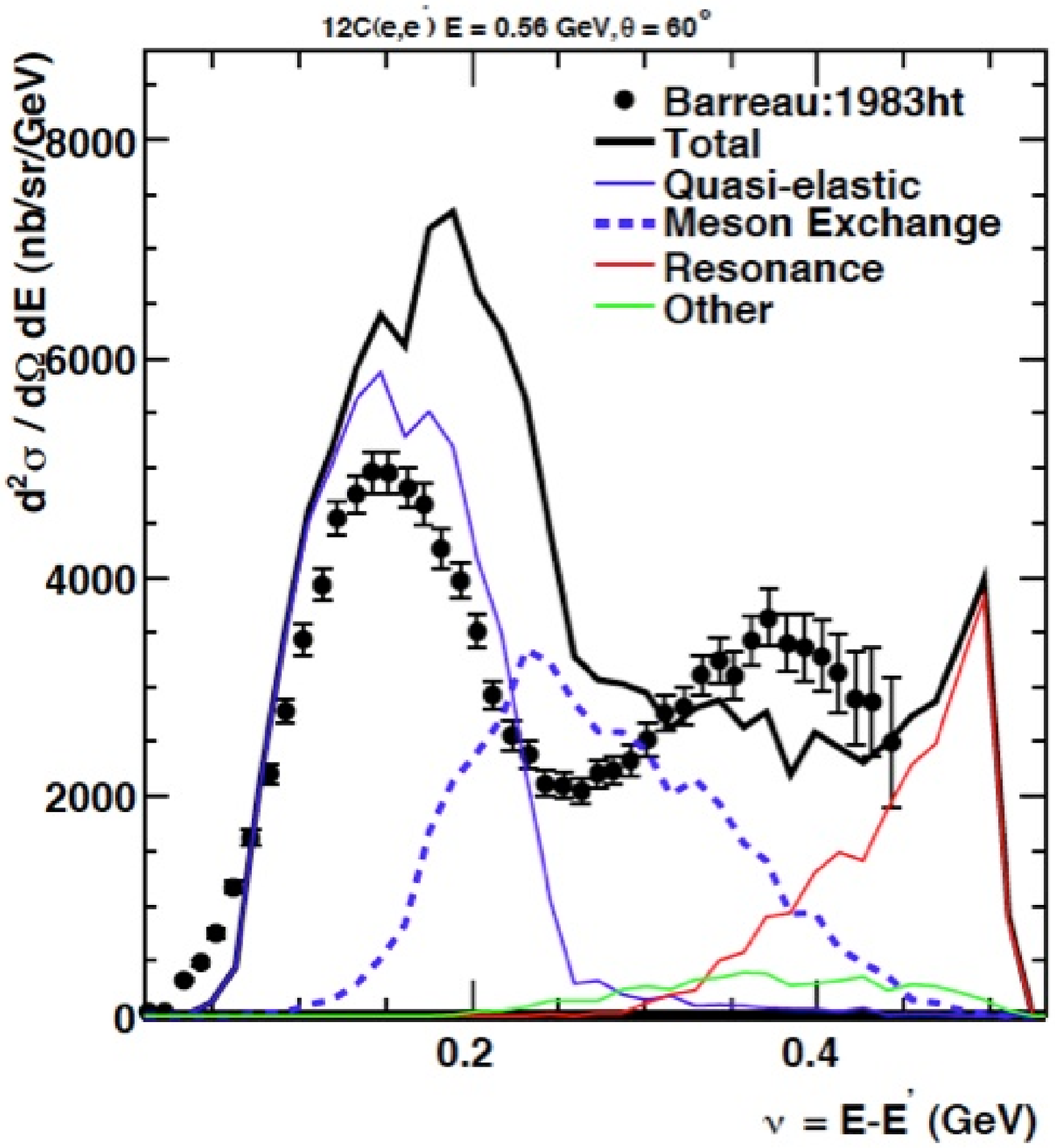}
\includegraphics[height=0.35\textheight]{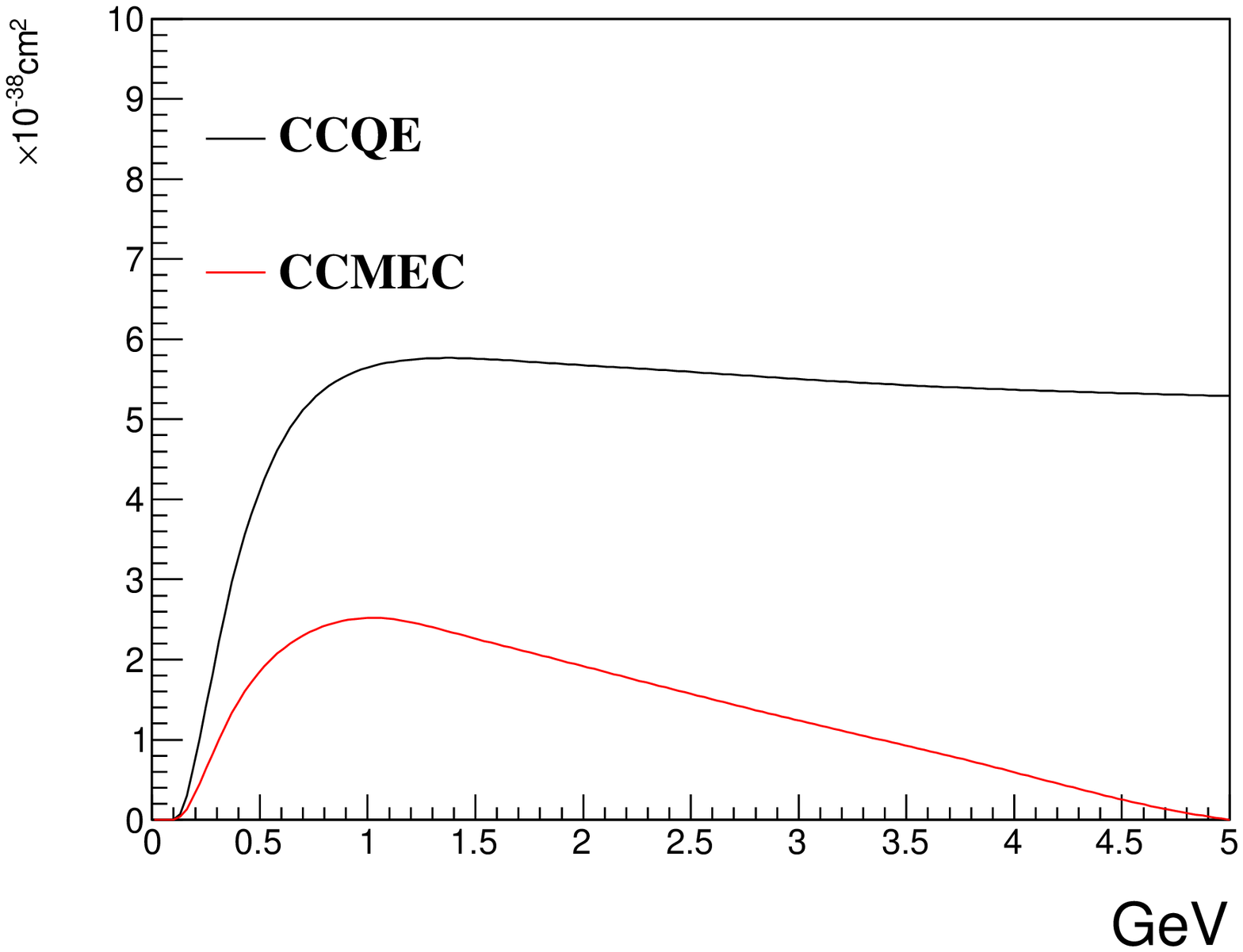}
\caption{\label{fig:dytman}
Dytman model predictions. 
The left plot shows data-simulation comparison of the 560~MeV electron inclusive cross section 
with 60 degree scattering. 
The MEC contribution can be seen between QE and resonance peaks.
The right plot shows energy dependence of the neutrino total cross section 
with a function of neutrino energy.}
\end{figure}

The goal is to tune the MEC strength from the electron scattering data, 
and apply the same model to the neutrino cross section prediction. 
The agreement is tested with electron scattering data. 
Figure~\ref{fig:dytman} (left) shows an example of inclusive cross section prediction. 
The data points are taken from Ref.~\cite{Barreau}. 
The model tuning using electron scattering data is ongoing.    
In this moment (GENIE v2.7.1), 
the strength is chosen so that it agrees with MiniBooNE and NOMAD data. 
Since NOMAD~\cite{NOMAD} does not observe the enhancement, 
the Dytman model is chosen to turn off monotonically 
from 1~GeV to 5~GeV (Fig.~\ref{fig:dytman}, right).

\subsection{Hadronic model, nucleon cluster model}

\begin{figure}[t]
\includegraphics[height=0.22\textheight]{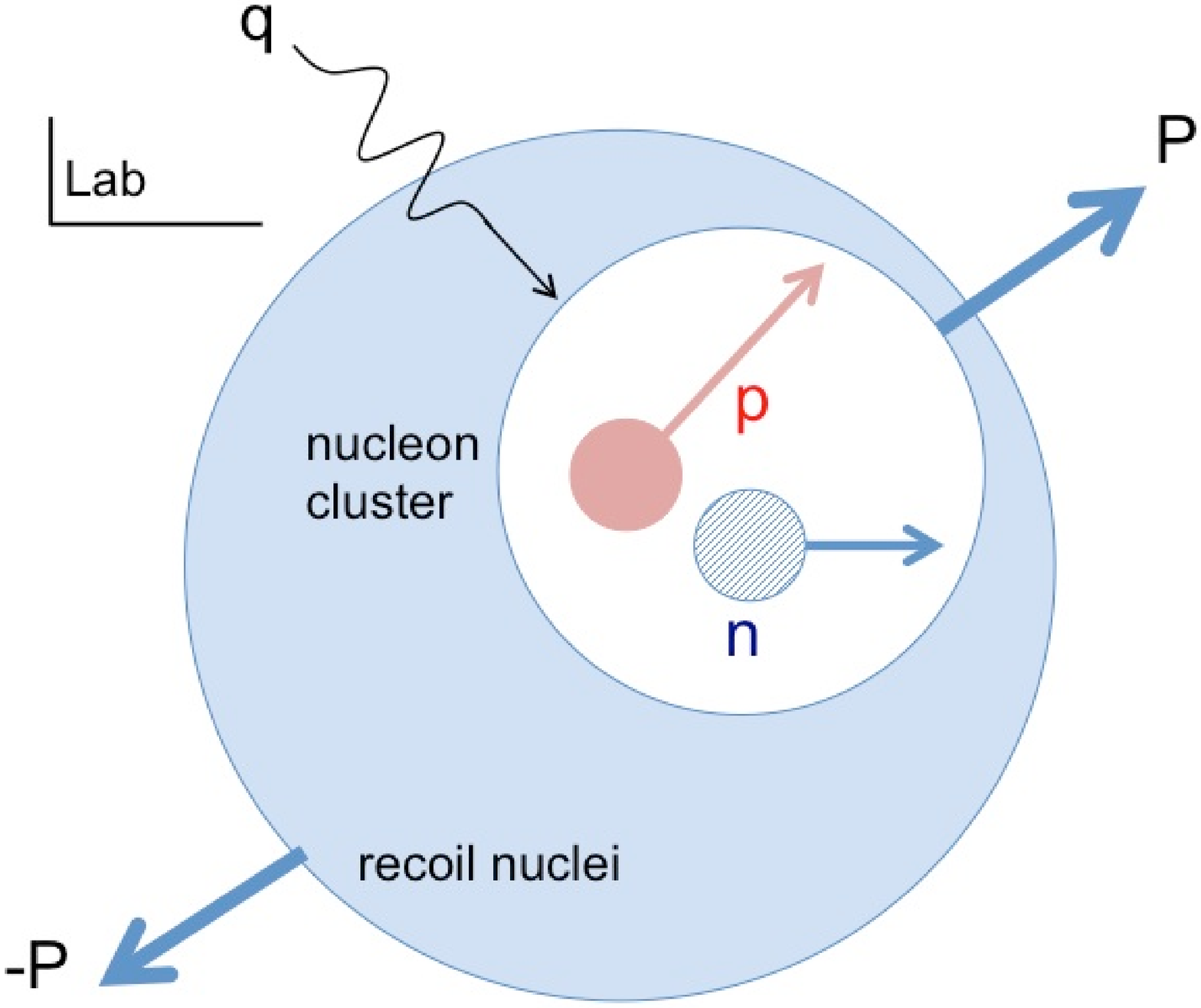}
\includegraphics[height=0.22\textheight]{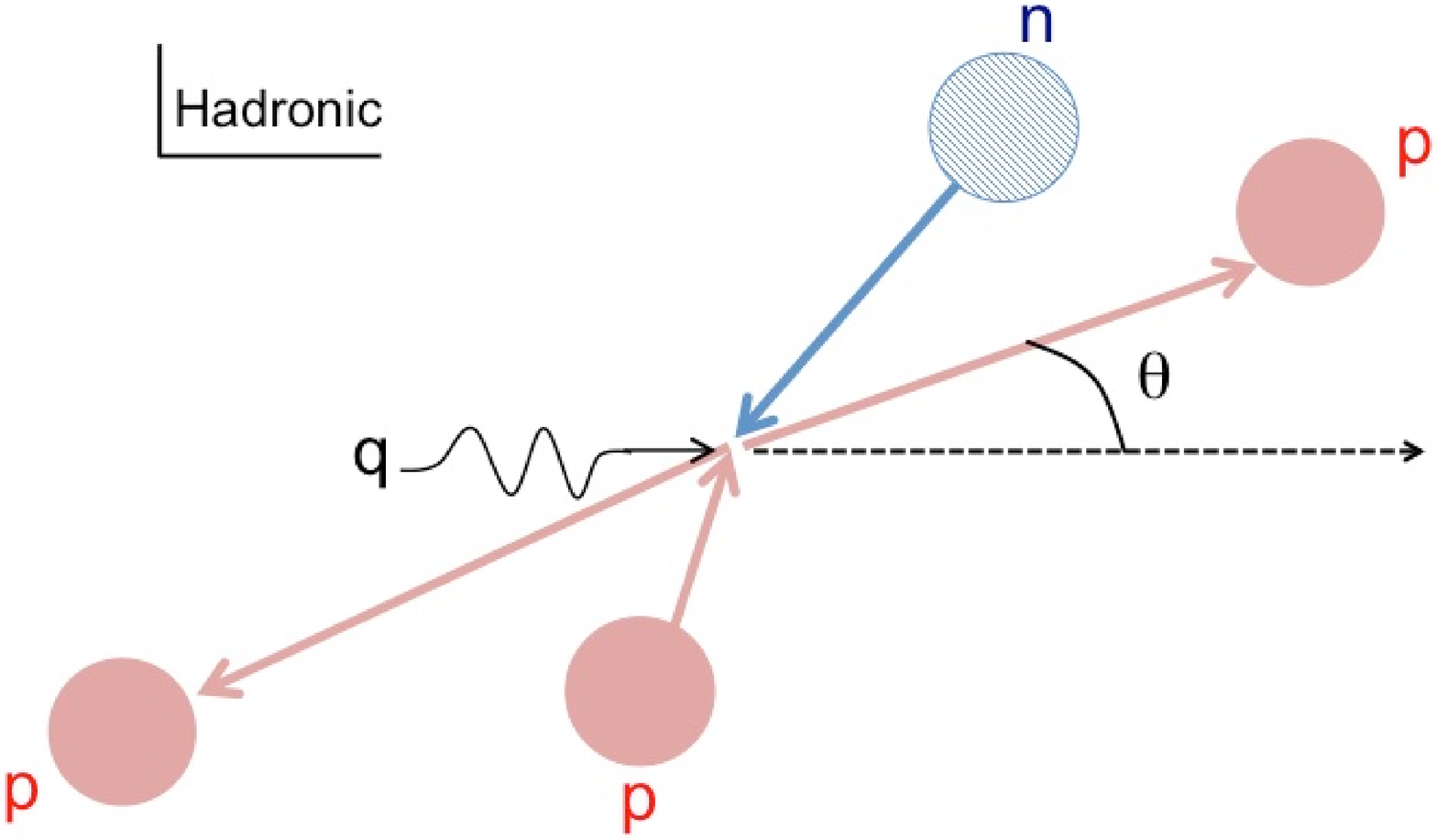}
\includegraphics[height=0.22\textheight]{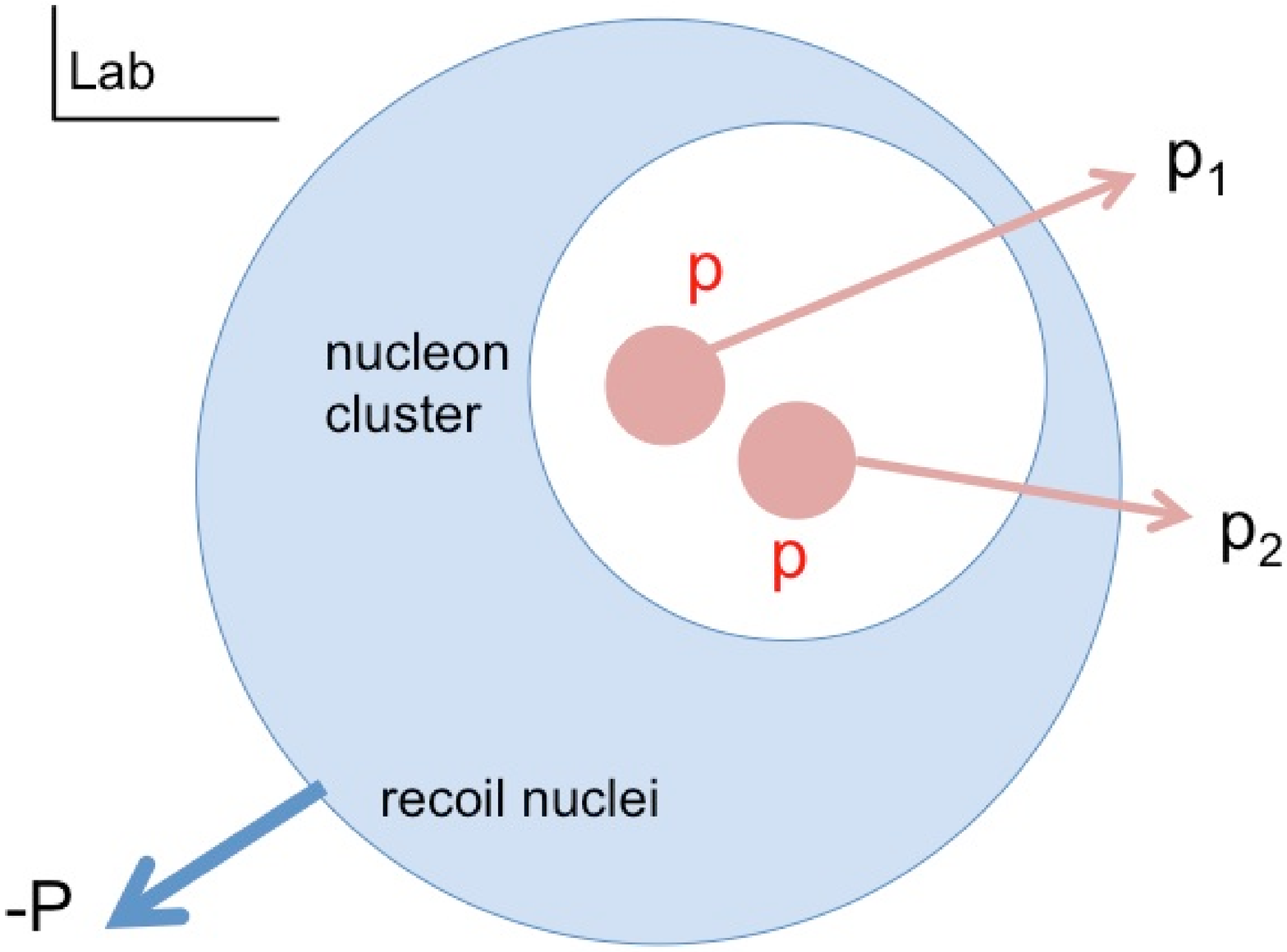}
\caption{\label{fig:cluster}
Nucleon cluster model in GENIE. First, 2 nucleons are chosen from the Fermi sea (left). 
Then 2 nucleons and energy-momentum transfer 4-vector make the hadronic system (middle). 
Finally, the system is boosted back, and 2 outgoing nucleons are generated in the lab frame (right).}
\end{figure}

For the hadronic model, the GENIE collaboration developed the nucleon cluster model. 
The hadronic model is important because this specifies outgoing nucleon types and kinematics. 
The nucleon cluster model is a naive approach, but reasonable 
(a similar model was independently developed in NuWro~\cite{nuwro_mec}). 
The procedure is depicted in Figure~\ref{fig:cluster}. 
Here 2 nucleons are independently selected from the Fermi sea, 
and the sum of them makes a ``nucleon cluster'' (Fig.~\ref{fig:cluster}, left) with momentum $P$ 
where the leftover nuclear recoils with momentum $-P$. 
Here, no Pauli blocking is applied. Also no separation energy is considered. 
Then, this nucleon cluster and energy-momentum transfer 
4-vector form the center-of-mass system, called the ``hadronic system'' (Fig.~\ref{fig:cluster}, middle), 
and isotropic decay in the hadronic system is boosted back to the lab frame 
to produce 2 outgoing nucleons (Fig.~\ref{fig:cluster}, right). 

\subsection{Final state interaction (FSI), hA model}

For the FSI model, the MEC model in GENIE uses the hA model as a default. 
More detail can be found elsewhere~\cite{dytman_nuint11}. 

\subsection{Results}

Here, simulation results are generated by folding with 
the MiniBooNE $\nu_\mu$ flux file~\cite{MB_ccqe}. 

\subsubsection{Lepton kinematics}
\begin{figure}[t]
\includegraphics[height=0.22\textheight]{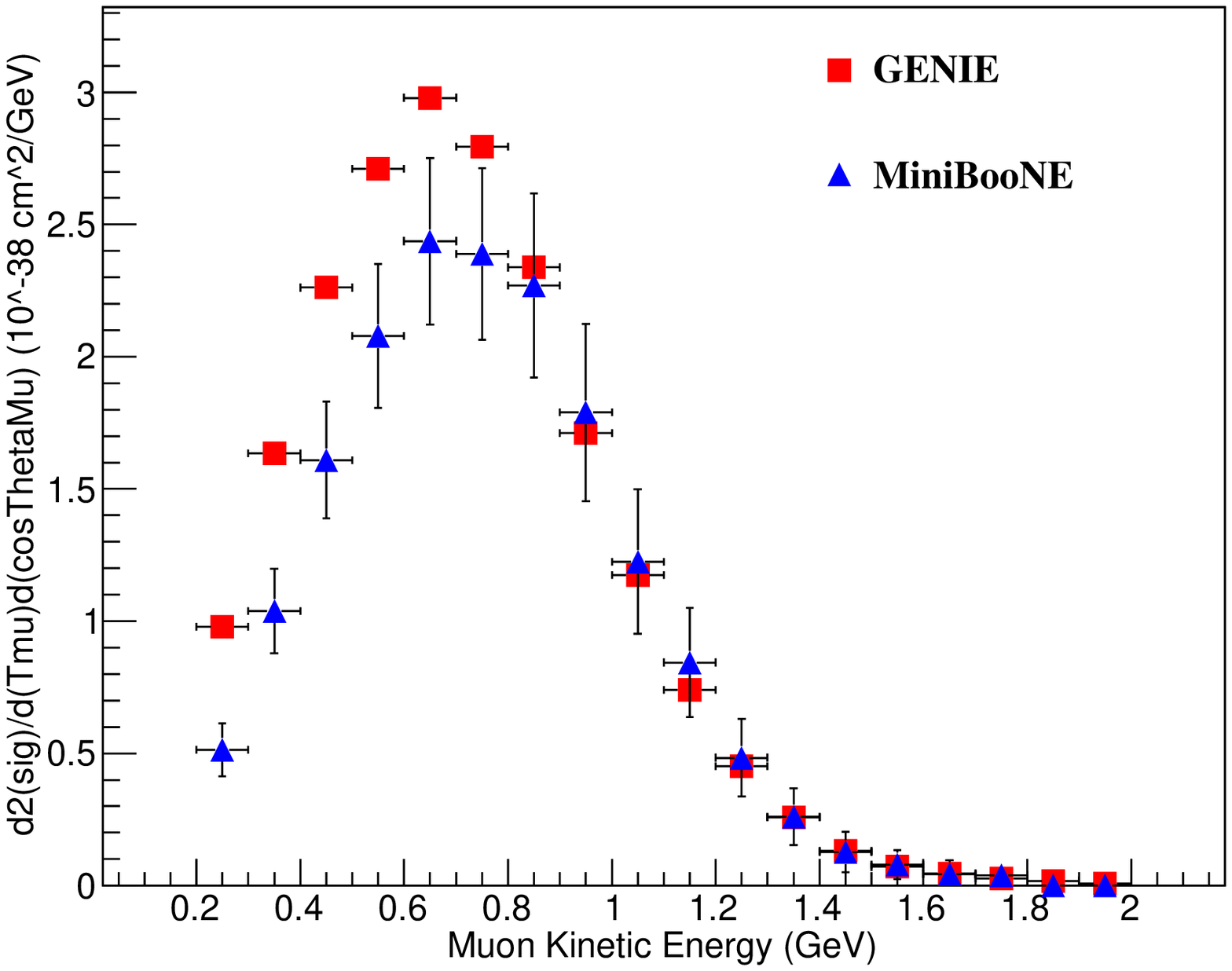}
\includegraphics[height=0.22\textheight]{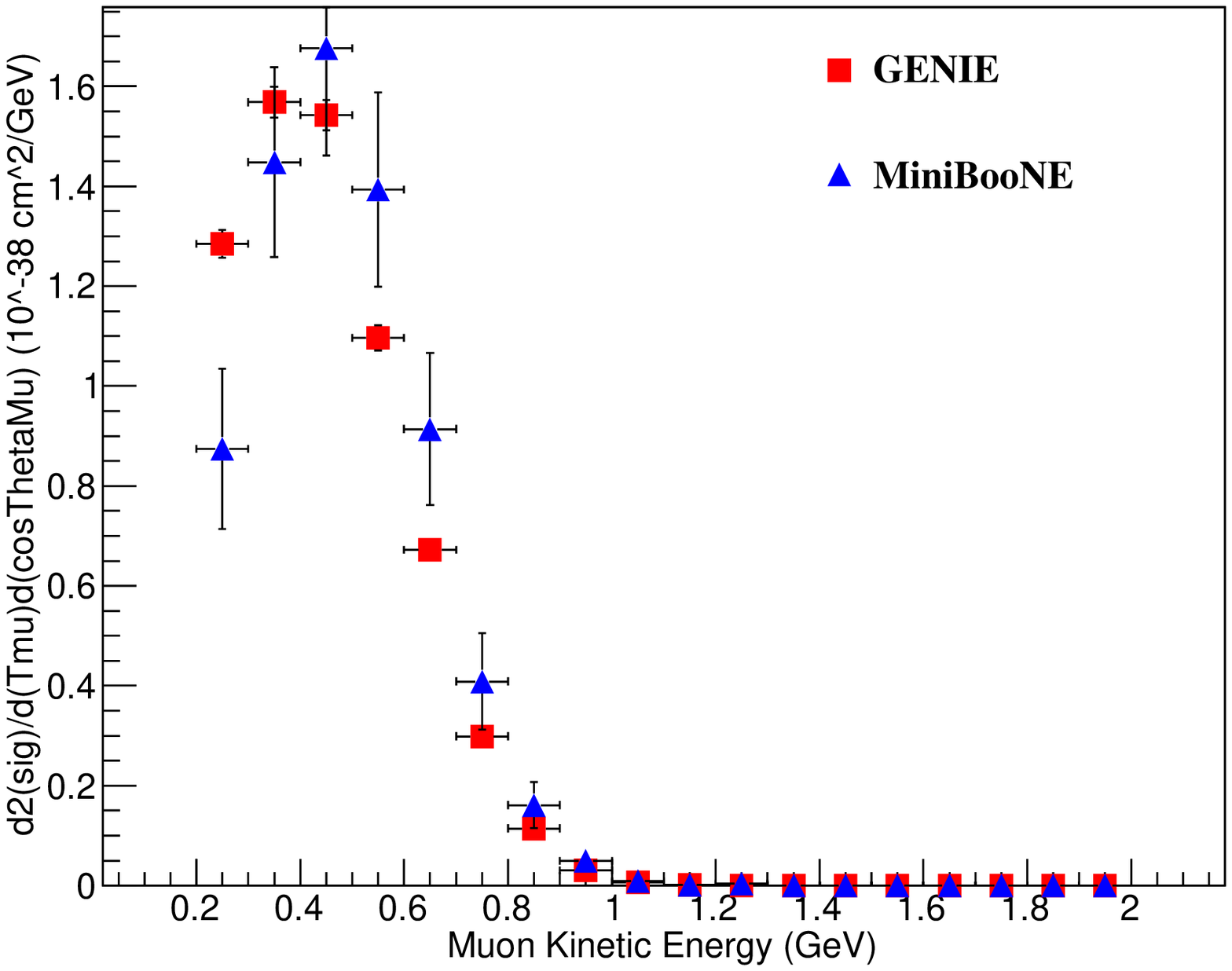}
\includegraphics[height=0.22\textheight]{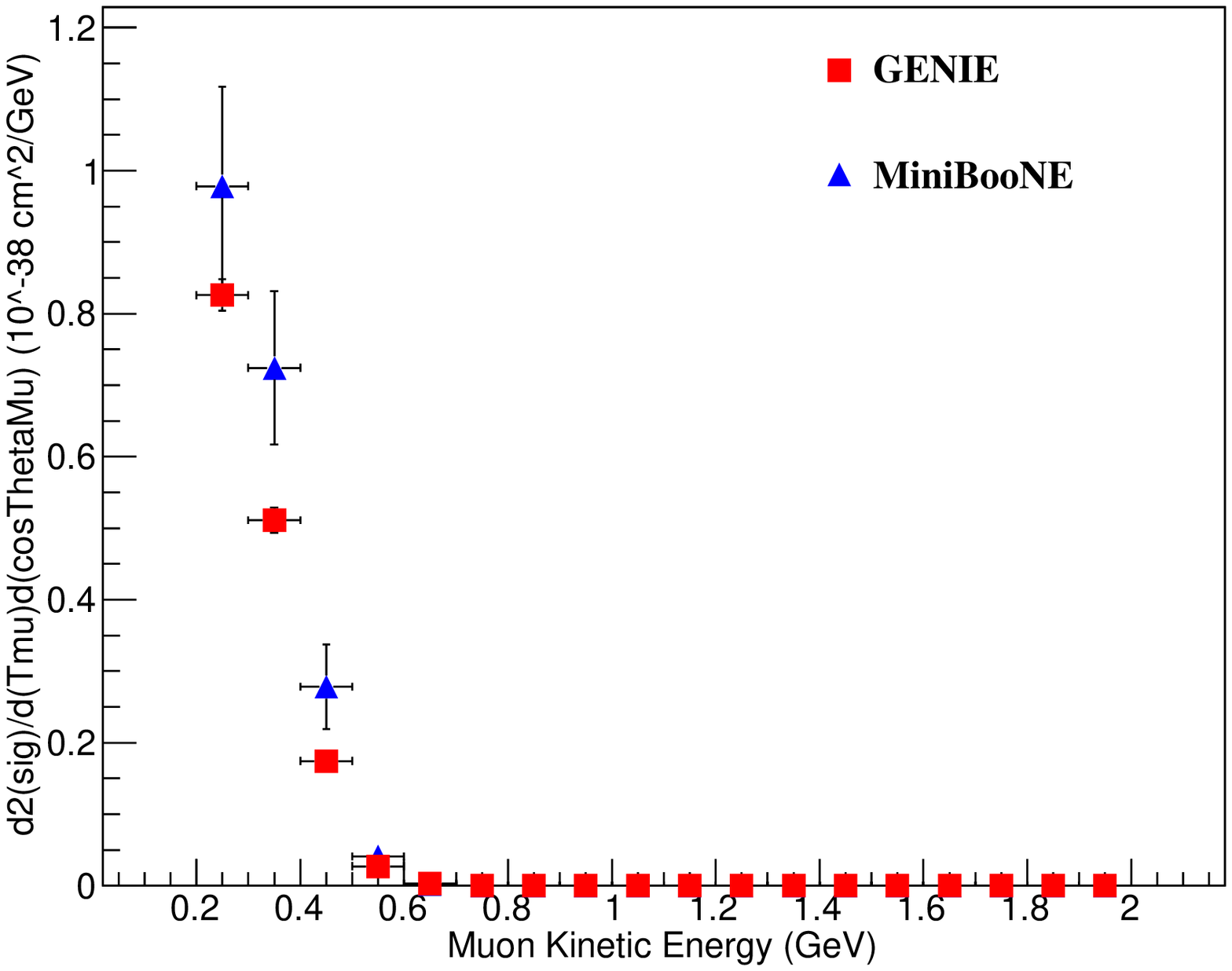}
\caption{\label{fig:mb_mu}
Comparisons of GENIE and MiniBooNE CCQE double differential cross section. 
Here all plots are a function of muon kinetic energy, in a slice of muon scattering angle.
CCQE is defined ``one muon, no pion, any number of nucleons'' in the detector. 
From left to right, $0.8<cos\th_\mu<0.9$, $0.5<cos\th_\mu<0.6$, $-0.1<cos\th_\mu<0.0$.}
\end{figure}

Figure~\ref{fig:mb_mu} shows the comparison of GENIE with 
the MiniBooNE CCQE double differential cross section data. 
In MiniBooNE, cross section channels are defined from the final state particles in the detector 
instead of interaction types at the primary neutrino interaction vertices~\cite{teppei}. 
Here MiniBooNE defines CCQE as 
``one muon, no pion, any number of nucleons'' in final state particles in the detector. 
The Dytman model has a good agreement with the MiniBooNE data at a moderate scattering angle, 
but it slightly overestimates at forward going muons, and underestimates back scattered muons. 
The tuning with larger set of electron scattering data is ongoing.

\subsubsection{Hadron kinematics}
\begin{figure}[t]
\includegraphics[height=0.22\textheight]{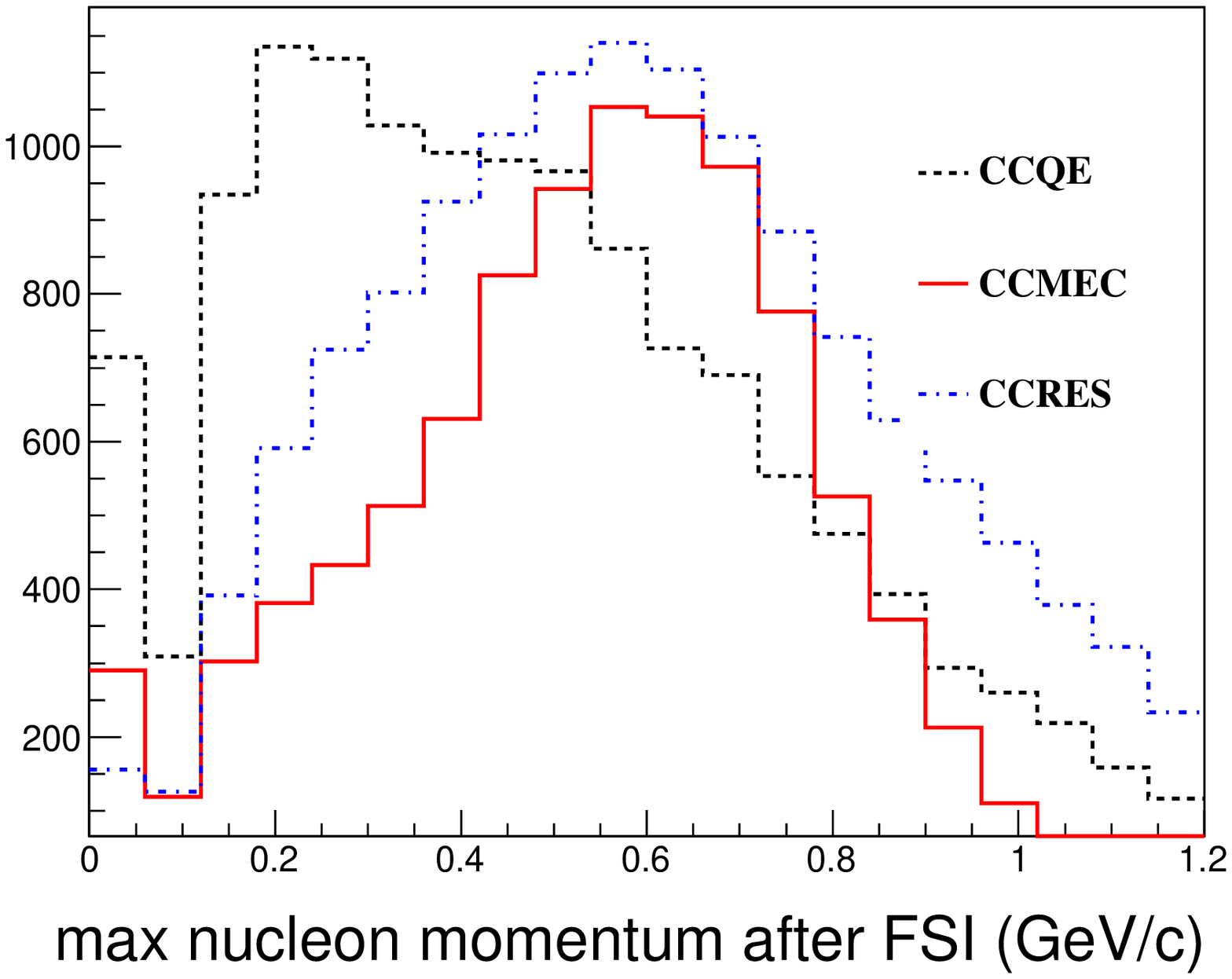}
\includegraphics[height=0.22\textheight]{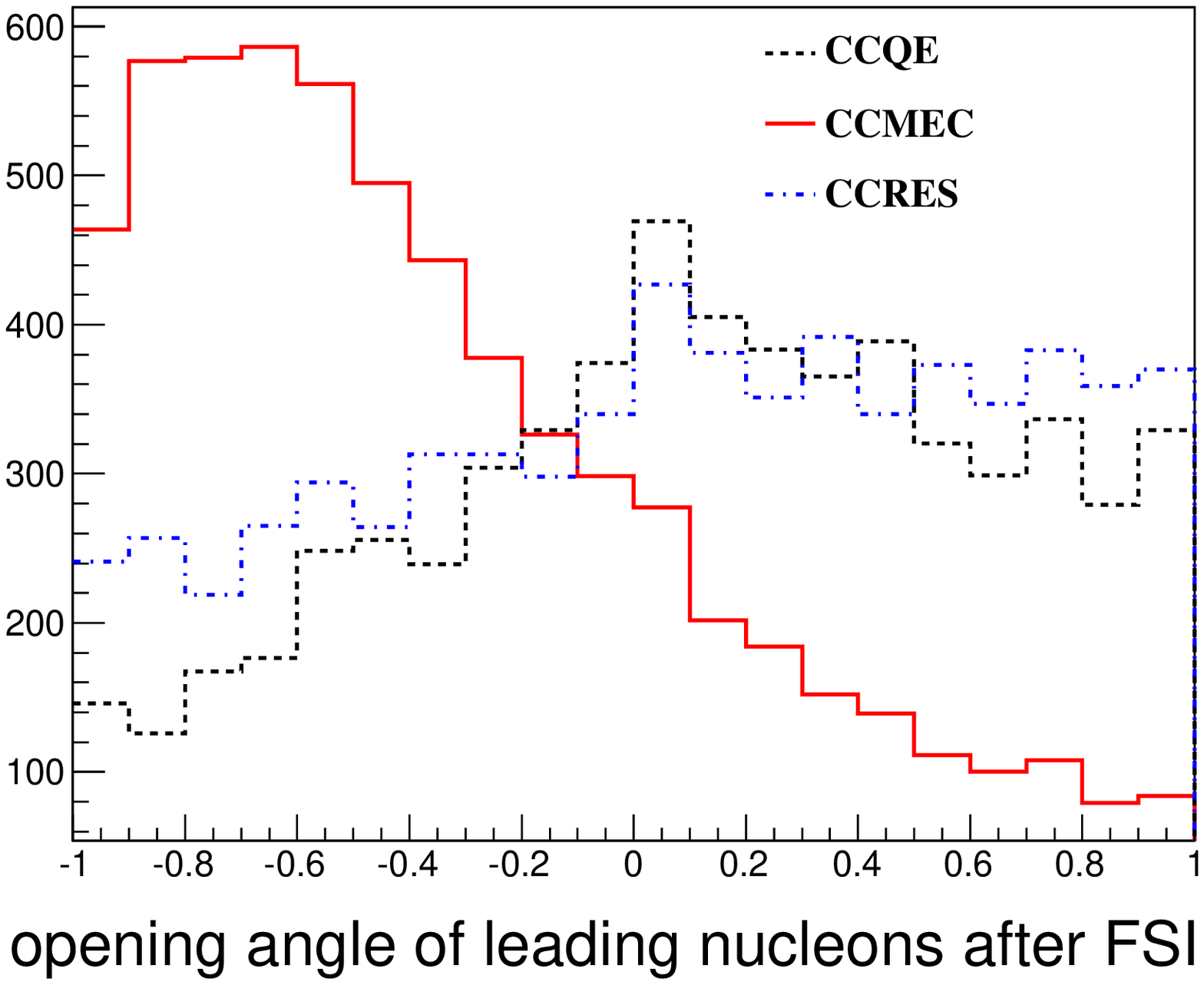}
\includegraphics[height=0.22\textheight]{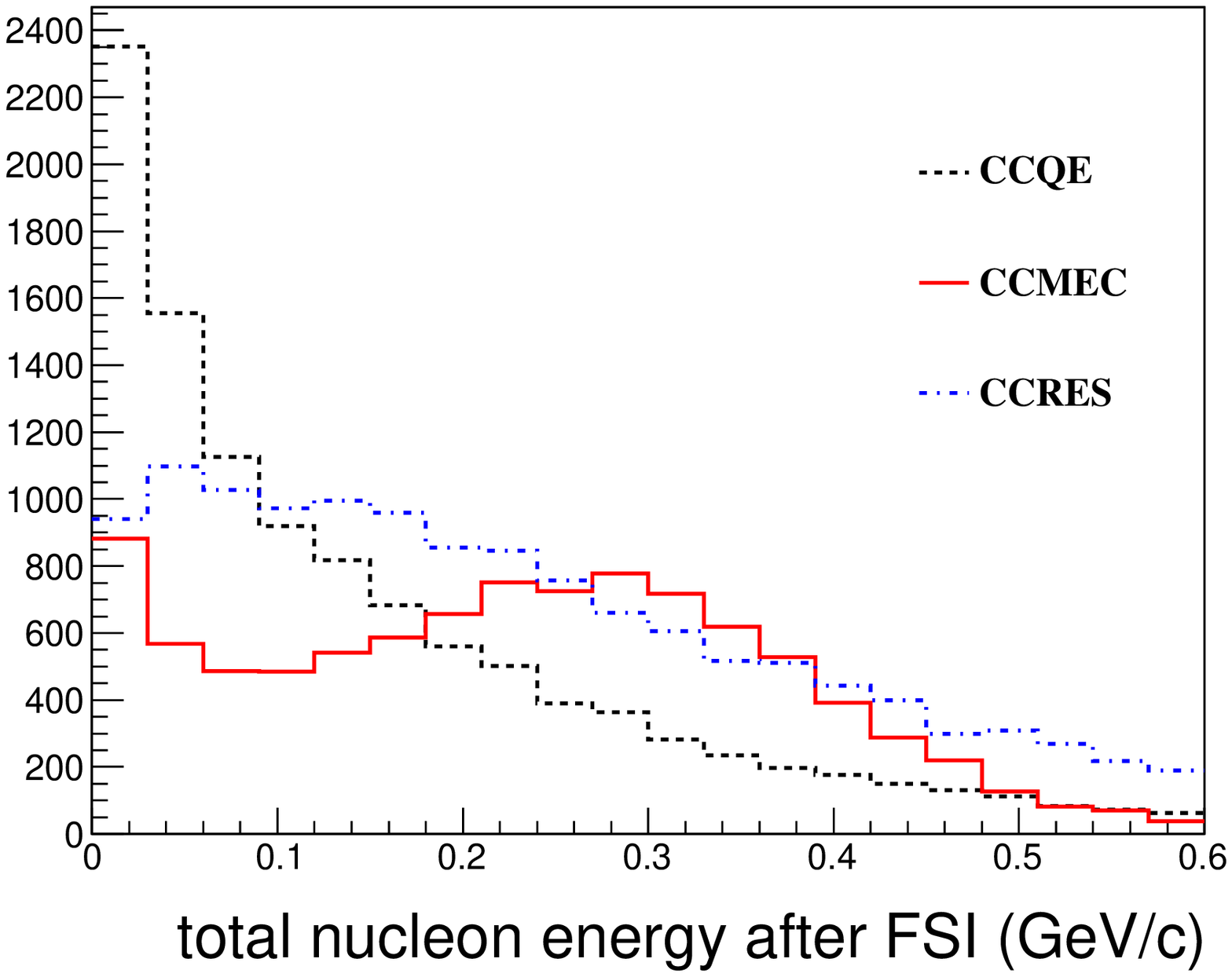}
\caption{\label{fig:mb_p}
Predictions of GENIE MEC model on nucleon kinematics. 
The left plot shows the momentum of the leading nucleon, 
the middle plot shows the opening angle of the 2 leading nucleons, 
and the right plot shows the sum of all nucleon kinetic energies. 
Note histograms are arbitrarily normalized to emphasize the shape differences.}
\end{figure}

It may be more interesting to take a look hadronic system prediction of GENIE. 
Figure~\ref{fig:mb_p} shows examples of the GENIE prediction 
of nucleon kinematic distributions. 
Here histograms are arbitrarily normalized to emphasize the shape differences. 
The left plot shows the leading nucleon momentum. 
Since MEC shares energy-momentum transfer with 2 nucleons, 
the maximum available nucleon momentum is lower than what simple CCQE predicts. 
This indicates that it may be a little harder to see the tracks of nucleons from the MEC. 
The middle plot shows the opening angle of the 2 leading nucleons. 
Although it is isotropic for the CCQE and resonance processes where 
the second leading nucleon is created by FSIs, 
MEC clearly predicts larger angles at the MiniBooNE neutrino beam energy ($\sim 800$~MeV) 
due to the nucleon cluster model.  
The right plot shows the sum of all nucleon kinetic energies. 
Such quantity may be more useful if experiments cannot identify multiple nucleon tracks~\cite{nuwro_mec}. 
Here, GENIE MEC model makes a characteristic peak around $\sim$0.3~GeV due to the Dytman model. 
This may be compared with predictions from other generators based on different leptonic models~\cite{nuwro_mec}. 

\section{Other neutrino interaction generators}

The model of the 2p-2h effect is also developed in NuWro~\cite{nuwro_general} 
and GiBUU~\cite{gibuu_general} neutrino interaction generators. 

\subsection{NuWro}
NuWro predictions of the nucleon kinematic distributions are shown in Ref.~\cite{nuwro_mec}. 
Hadronic kinematics predictions are related with not only the hadronic model, 
but also the leptonic model, 
because the leptonic model specifies the energy-momentum transfer given to the hadronic system 
to emit nucleons. 
Currently, NuWro prepares 3 different leptonic models, 
np-nh model based on Marteau {\it et al.}~\cite{nuwro_marteau}, 
Valencia model from Nieves {\it et al.}~\cite{nieves_first}, 
and Bodek {\it et al.}'s transverse enhancement model (TEM)~\cite{bodek_tem}. 

\subsection{GiBUU}
GiBUU~\cite{mosel_dd} has more emphasis on the FSI part, 
by taking account the phase space evolution. 
Interestingly, GiBUU predicts no MEC contribution in 
``one muon, zero pion, zero proton'' CC sample. 
GENIE agrees with this result. 
This opens a new way to measure genuine CCQE interactions, 
since precise counting of outgoing nucleons may nail down the primary process. 
NOMAD~\cite{NOMAD}, K2K~\cite{K2K}, 
and SciBooNE~\cite{SB_ccpip} showed the zero and one proton counting from their CC samples, 
but future high resolution experiments, such as MicroBooNE~\cite{georgia}, 
can push this idea further.  

\section{Conclusion} 

Table~\ref{tab:generator} shows a comparison of MEC models in various generators. 
Clearly, the hadronic model is the key to find the MEC processes in the neutrino scattering program.  
Are there any correlations in initial nucleon kinematics? 
What kind of pairs are favored, proton-neutron pairs or neutron-neutron pairs? and by how much? 
How to share the energy-momentum transfer between 2 nucleons? etc. 
All these are open questions!
Theorists are encouraged to provide this information to experimentalists.  

\begin{table}
\begin{tabular}{lrrr}
\hline
  & \tablehead{1}{r}{b}{GENIE}
  & \tablehead{1}{r}{b}{NuWro}
  & \tablehead{1}{r}{b}{GiBUU}  \\
\hline
Leptonic model & Dytman model & TEM, np-nh model, & Transverse projector\\
               &              & and Valencia model & \\
Hadronic model & nucleon cluster & nucleon cluster & phase space density\\
initial nucleon momentum & Fermi sea & Fermi sea & Fermi sea\\
initial nucleon momentum correlation&none&none&none\\
initial nucleon spatial correlation &none&none&2 nucleons are generated\\
                                   &    &    &at the same location\\
initial nucleon pair& n-p:n-n=1:4& n-p:n-n=9:1&n-p:n-n=12:5\\
                   &isospin ansatz&short range correlation&statistical average\\
FSI model& hA model& cascade model & BUU transport \\
\hline
\end{tabular}
\caption{Comparison of MEC models in neutrino interaction generators.}
\label{tab:generator}
\end{table}

The GENIE collaboration is preparing to release the next frozen version, v.2.8.0. 
Here, the MEC model described in this paper will be included. 
There is an ongoing effort for the further tuning 
of this channel based on the larger set of electron scattering data. 
 
\begin{theacknowledgments}
The author thanks the organizer for the invitation to the conference, 
and the hospitality during my stay in Rio de Janeiro. 
The author thanks Costas Andreopoulos, Steve Dytman, 
and Hugh Gallagher of GENIE collaboration for providing information, 
and Jan Sobczyk and Ulrich Mosel for useful discussions, 
and Brandon Eberly for sharing a code. 
The author also thanks Ben Jones and Jennifer Dickson for the careful reading of this manuscript.
\end{theacknowledgments}
\bibliographystyle{aipproc}   

\bibliography{sample}

\end{document}